\documentclass[12pt]{article}
\usepackage{amsmath,amssymb,graphicx,slashbox,epsfig,color}

\def\mydate{August 3, 2009}
\def\ignore#1{{}}

\ignore{
\setlength{\textwidth}{150mm} 
\setlength{\textheight}{230mm} 
\setlength{\oddsidemargin}{5mm} 
\setlength{\evensidemargin}{5mm} 
\setlength{\topmargin}{-5mm} 
\makeatletter
  
  \@addtoreset{equation}{section}
\makeatother

}

\renewcommand{\baselinestretch}{1.25}

\renewcommand{\thefootnote}{\arabic{footnote}}

\newcommand{\beeq}{\begin{equation}}
\newcommand{\eneq}{\end{equation}}
\newcommand{\beqn}{\begin{eqnarray}}
\newcommand{\eeqn}{\end{eqnarray}}

\def\dd{\partial}
\def\la{\raise.16ex\hbox{$\langle$}\lower.16ex\hbox{}  }
\def\ra{\raise.16ex\hbox{$\rangle$}\lower.16ex\hbox{} }
\def\go{\rightarrow}

\def\onehalf{ \hbox{${1\over 2}$} }

\def\Tr{{\rm Tr \,}}

\def\eff{{\rm eff}}

\def\diag{{\rm diag ~}}

\def\KK{{\rm KK}}

\def\psibar{ \psi \kern-.65em\raise.6em\hbox{$-$} }
\def\psibarl{ \psi \kern-.65em\raise.6em\hbox{$-$} \lower.6em\hbox{} }

\begin{document}

\thispagestyle{empty}

\baselineskip=12pt

{\small \noindent \mydate    \hfill OU-HET 631/2009}

{\small \noindent                 \hfill KIAS-P09040}


\baselineskip=35pt plus 1pt minus 1pt

\vskip 2.5cm

\begin{center}
{\LARGE \bf Stable Higgs Bosons as Cold Dark Matter}\\

\vspace{2.5cm}
\baselineskip=20pt plus 1pt minus 1pt

{\def\thefootnote{\fnsymbol{footnote}}
\bf 
Yutaka Hosotani,$^*$  Pyungwon Ko$^\dagger$ and Minoru Tanaka$^*$}\\

\vspace{.3cm}
{\small  $^*${\it Department of Physics, Osaka University,
Toyonaka, Osaka 560-0043, Japan}}\\
{\small  $^\dagger${\it School of Physics, KIAS,
Seoul 130-722, Korea}}\\
\end{center}

\vskip 2.0cm
\baselineskip=20pt plus 1pt minus 1pt

\begin{abstract}
In a class of the gauge-Higgs unification models the 4D neutral Higgs boson,
which is a part of the extra-dimensional component of the gauge fields, 
becomes absolutely stable as a consequence of the gauge invariance and dynamically generated new parity, 
serving  as a promising candidate for cold dark matter (CDM).  
We show that the observed relic abundance of 
cold dark matter
is obtained in the $SO(5) \times U(1)$ model in the warped space
with the Higgs mass  around 70$\,$GeV.
The Higgs-nucleon scattering cross section is found to be
close to the current CDMS II and XENON10 bounds 
in the direct detection of dark matter.
\end{abstract}


\newpage

What constitutes dark matter in the universe?$\,$\cite{DM1}-\cite{BHS}
How does the Higgs boson  interact with other particles?$\,$
\cite{djouadi}
These are two of the most important issues in current physics.  
We would like to point out
that these two issues are related to each other, 
and indeed that they reflect two sides of the fact that 
Higgs bosons can be stable and become dark matter.

In the standard model (SM) of electroweak interactions, the electroweak 
(EW)  symmetry is spontaneously broken by the Higgs field.
However it is not clear at all   if the Higgs boson appears as envisaged 
in the standard model.  
It is often argued  that the naturalness and stability against radiative
corrections to the mass of the Higgs field indicate the existence of
supersymmetry underlying the nature.
However, there are other  scenarios with the naturalness. 
The gauge-Higgs unification scenario is one of them, in which 
the 4D Higgs field, $H(x)$, is a part of the extra-dimensional component
of the gauge fields. It is identified as the fluctuation
mode of the Aharonov-Bohm  phase (Wilson line phase), $\theta_H$, 
 in the extra-dimension.\cite{YH1}-\cite{YH2}

It has been recently found  in one  of the  gauge-Higgs 
unification models in the Randall-Sundrum warped space\cite{RS} that 
the Higgs mass $m_H$ comes out around 50 GeV and the Higgs couplings
with $W$ and $Z$ bosons,
$WWH$ and $ZZH$, and Yukawa couplings with fermions,  $H \psibar \psi$, vanish
at the value of $\theta_H = \pm \onehalf \pi$ where the effective potential
for $\theta_H$ is minimized.\cite{HOOS, HK}  If this is the case,  
the Higgs boson would become stable.

This prompts urgent questions.  
Does  the  Higgs couplings 
to the SM fermions and weak gauge bosons  
vanish to all order in perturbation theory?  Is the value 
$\theta_H = \pm \onehalf \pi$
naturally achieved?  Can  the Higgs boson be absolutely stable?
What  physical  consequences follow from stable Higgs bosons?

We show in this paper that  the  $WWH$, $ZZH$,  and $H \psibar \psi$
couplings exactly vanish and the Higgs boson becomes stable, provided that
certain conditions for the matter content are satisfied in the
$SO(5) \times U(1)$ gauge-Higgs unification in the Randall-Sundrum
warped space. 
\ignore{Higgs bosons are produced in pairs. 
As they are stable, they appear in collider experiments as missing
energies and momenta. The way of searching for Higgs bosons in collider
experiments must be altered as well.}

This leads to an  important consequence. Higgs bosons become viable
candidates for cold dark matter (CDM) in the universe.  
They are copiously produced in the very early universe.  
As the annihilation rate of Higgs bosons falls below the expansion rate 
of the universe,  the annihilation processes get frozen
and the remnant Higgs bosons become dark matter. 

The annihilation rates of Higgs bosons can be
reliably estimated, once the masses of $W$, $Z$, and fermions are given.
The precise value of the Higgs mass $m_H$, which arises as quantum 
effects of $\theta_H$, depends on the details of the model.  
In the model of ref.\ \cite{HOOS},  $m_H$ is predicted to be 
$\sim 50$ GeV.  
It is shown below that the observed cold dark matter density is obtained
with $m_H \sim 70 $ GeV in the $SO(5) \times U(1)$ model. 

The distinctive feature  in the gauge-Higgs unification is 
that the Higgs field $H(x)$ corresponds to the fluctuation of
the Aharonov-Bohm (AB) phase, $\theta_H$,   in the extra-dimensional space,
which is given by the phase of the path-ordered Wilson line integral along
a noncontractible loop $C$ in the extra-dimension with a coordinate $y$, 
$P \exp  \big\{i g_A \int_C dy A_y(x,y) \big\}$.
The effective Higgs interactions with the 
$W$, $Z$ bosons and fermions in the standard model at low energies are
summarized in the $SO(5) \times U(1)$ model as \cite{HK}
\beqn 
&&\hskip -1cm
{\cal L}_\eff  = - V_\eff (\hat \theta_H) 
                 - m_W^2(\hat \theta_H) W^\dagger_\mu W^\mu
                 - \onehalf m_Z^2(\hat \theta_H) Z_\mu Z^\mu 
                 - \sum_{a,b} m^F_{ab}(\hat \theta_H) \psibar_a  \psi_b ~, \cr
\noalign{\kern 5pt} 
&&\hskip -1 cm
\hat \theta_H (x) = \theta_H + \frac{H(x)}{f_H}
~~,~~
f_H = \frac{2}{g_A} \sqrt{\frac{k}{z_L^2 -1}}
\sim \frac{2}{\sqrt{kL}} \frac{m_\KK}{\pi g} ~~. 
\label{effective1}
\eeqn
Here $k$ and $L$ are two parameters specifying the Randall-Sundrum warped
space.\cite{RS}  
The fundamental region in the fifth coordinate is  $0 \le y \le L$. 
The AdS curvature in the bulk region is given by  $\Lambda = - 6 k^2$.  
The warp factor $z_L = e^{kL}$ and the Kaluza-Klein (KK) mass scale  
$m_\KK = \pi k z_L^{-1}$ are, typically, $10^{15}$ and $1.5\,$TeV, respectively.
The $SO(5)$ gauge coupling $g_A$ is related to the 4D $SU(2)_L$ weak 
coupling $g$ by $g=g_A/\sqrt{L}$.  

The effective potential $V_\eff (\hat \theta_H)$ is dynamically generated at the one-loop
level, and is finite.\cite{YH1}  The mass functions $m_W(\hat \theta_H)$, 
$m_Z(\hat \theta_H)$, and $m^F_{ab}(\hat \theta_H)$ appear at the tree level,
which include contributions from heavy KK 
excited states in intermediate states.\cite{HK, HS2, Sakamura}
The value of the AB phase $\theta_H$ is dynamically determined by
the location of the global minimum of $V_\eff (\theta_H)$.  
There  the fermion fields are  diagonalized in the mass basis such that 
$m^F_{ab} = m^F_a \delta_{ab}$.

In the $SO(5) \times U(1)$ model in the RS warped space
\cite{Agashe1, HS2, MSW} the orbifold boundary conditions 
\beeq
\begin{pmatrix} A_\mu \cr - A_y \end{pmatrix} (x, y_j - y) =
P_j \begin{pmatrix} A_\mu \cr  A_y \end{pmatrix} (x, y_j + y) P_j^{-1}
~, ~~ (j=0,1),
\label{BC1}
\eneq
are imposed for $SO(5)$ gauge fields,  where  $(y_0 , y_1)=(0, L)$.  
Here $P_j$'s are elements of $SO(5)$.  With the parity matrices 
$P_j = P_j^{\rm vec}  = \diag (-1,-1,-1,-1,1)$ in the vectorial representation,
$SO(5)$ is broken to $SO(4) \simeq SU(2)_L \times SU(2)_R$.  
The 4D Higgs fields $\phi_a(x)$ ($a= 1, \cdots, 4$)
appear as zero modes in the $SO(5)/SO(4)$ part of $A_y(x,y)$.
In components 
$A_y^{\hat a}(x,y)  = \phi_a(x) h_0(y) + \cdots $ 
where $h_0(y)$ is the zero mode wave function  given below.
An $SO(4)$ vector $\phi_a$ forms an $SU(2)_L$ doublet corresponding
to the Higgs doublet in the standard model.  $\phi_1, \phi_2, \phi_3$ are
absorbed by $W$ and $Z$, whereas  $\phi_4$ corresponds to the physical
neutral Higgs field.

Let us denote the generators of $SO(5)/SO(4)$ by 
$T^{\hat{a}}$ $ (a=1,\cdots,4)$. In the vectorial representation
$(T^{\hat 4}_{\rm vec})_{ab} =( i /\sqrt{2}) 
 (\delta_{a5} \delta_{b4} -\delta_{a4} \delta_{b5}) $,
whereas in the spinorial representation 
$T^{\hat 4}_{\rm sp} = ( 1 /2\sqrt{2}) I_2 \otimes \tau_1$.\cite{Agashe1}
The difference in normalization  
$\Tr T^{\hat a}_{\rm vec} T^{\hat b}_{\rm vec} = 
2 \, \Tr  T^{\hat a}_{\rm sp} T^{\hat b}_{\rm sp} = \delta_{ab} $ becomes
important in the subsequent discussions.

The fifth dimension in the Randall-Sundrum warped space has topology of 
the orbifold $S^1/Z_2$.  Its metric  satisfies
$g_{MN}(x,y+ 2L) = g_{MN}(x, y) = g_{MN}(x,-y)$.
Consequently the AB phase $\theta_H$ along the fifth dimension  is given by 
$\exp\{\frac{i}{2}\theta_H (2\sqrt2 \, T^{\hat{4}})\}
=\exp\{ ig_A \int^{L}_{0} dy  \la A_y\ra \}$.  
The relevant part of the gauge potential becomes 
\beeq
A_y (x,y) = \hat \theta_H (x) \cdot
 \sqrt{ \frac{4k}{z_L^2 -1} } ~ h_0(y)  \cdot
T^{\hat 4} + \cdots ~, 
\label{Ay1}
\eneq
where $h_0(y) = [2k/(z_L^2 -1)]^{1/2} \, e^{2ky}$ ($0 \le y \le L$) and
$\hat \theta_H(x)$ is defined in (\ref{effective1}).
$h_0(y)$ is normalized by $\int_0^L dy \, e^{-2ky} h_0(y)^2 = 1$, and
is given, outside the fundamental region,  by  $h_0(-y) = h_0(y) = h_0(y+2L)$.

$\theta_H$ is a phase variable.  
All the functions of $\hat \theta_H$  in the effective 
interaction in (\ref{effective1}) are periodic  with a period $2 \pi$.  
This periodicity follows from the large gauge invariance.\cite{YH2, HHHK, HM}  
Given the boundary conditions (\ref{BC1}), there remains the residual gauge invariance
$A_M \go A_M' = \Omega A_M \Omega^{-1} - (i/g_A) \Omega \dd_M \Omega^{-1}$
which preserves the boundary conditions.  In general, 
new gauge potentials satisfy new boundary conditions
\beeq
\begin{pmatrix} A_\mu' \cr - A_y' \end{pmatrix} (x, y_j - y) =
P_j'  A_M'  (x, y_j + y) {P_j'}^{-1}    
- \frac{i}{g_A} \, P_j'  \, \dd_M  {P_j'}^{-1}    ~, 
\label{BC2}
\eneq
where ${P_j'} = \Omega(x, y_j - y) P_j \Omega(x, y_j + y)^{-1}$. 
The residual gauge invariance is defined with $\Omega(x,y)$ satisfying
$P_j' = P_j$.

Consider a large gauge transformation 
$\Omega^{\rm large}(y) = \exp \big\{  i\alpha \int_0^y dy \,
\sqrt{4k/(z_L^2-1)} ~ h_0(y) \cdot T^{\hat 4} \big\}$
in the spinorial representation in which $P_j^{\rm sp} = I_2 \otimes \tau_3$. 
It shifts $\hat \theta_H(x)$ to $\hat \theta_H' (x) = \hat \theta_H(x) - \alpha$. 
It is straightforward to see that the condition ${P_j' }^{\rm sp} = {P_j}^{\rm sp}$ 
is satisfied  if $\alpha = 2\pi n$ ($n$: an integer).
In other words all physical quantities must be periodic in $\theta_H$ with a period
$2\pi$.  

Fermions in the vector representation of $SO(5)$, for instance,  obey
$\Psi (x, y_j - y) = {P_j}^{\rm vec} \gamma^5 \Psi(x, y_j + y)$.  
Under a gauge transformation  $\Omega^{\rm large}(y)$ with $\alpha = \pi$,
one finds that $({P_0'}^{\rm vec}, {P_1'}^{\rm vec} ) = ({P_0}^{\rm vec}, {P_1}^{\rm vec})$
whereas $({P_0'}^{\rm sp}, {P_1'}^{\rm sp} ) = ({P_0}^{\rm sp}, -{P_1}^{\rm sp})$.
Hence, if there are no fermions in the spinor representation of $SO(5)$, 
there appears enhanced symmetry.  In this case all physical quantities become
periodic in $\theta_H$ with a reduced period $\pi$.   
In the model of ref.\ \cite{HOOS}
bulk fermions appear only in the vector representation,  thereby  this condition 
being satisfied.
Brane fermions located at $y=0$ are not affected by the transformation 
as $\Omega^{\rm large}(0)= 1$.

There is mirror reflection symmetry in the extra dimension.  The action
in the RS warped spacetime  is invariant under 
$(x^\mu, y) \go (x'^\mu, y') = (x^\mu, -y)$, 
$A_M (x, y) \go A'_M (x', y') = (A_\mu, - A_y)(x, y)$,  and
$\Psi(x, y) \go \Psi' (x', y') = \pm \gamma^5 \Psi(x, y)$.
The orbifold boundary conditions are preserved under this transformation.
Since $h_0(-y) = h_0(y)$, this implies that the theory at low energies 
is invariant under $\hat \theta_H (x) \go \hat \theta_H' (x') = - \hat \theta_H (x) $.

In a class of the  $SO(5) \times U(1)$ gauge-Higgs unification models
in the warped space which contains fermions only in tensorial representations, 
but not in  spinorial representations, of $SO(5)$, one can draw an important
conclusion about the couplings of Higgs bosons.  
It follows from the enhanced gauge symmetry and
mirror reflection symmetry that
\beqn
&&\hskip -1cm
V_\eff (\hat \theta_H + \pi) = V_\eff (\hat \theta_H) 
 = V_\eff (- \hat \theta_H) ~, \cr
\noalign{\kern 5pt}
&&\hskip -1cm
m_{W, Z}^2(\hat \theta_H + \pi) = m_{W, Z}^2 (\hat \theta_H) 
= m_{W, Z}^2 (- \hat \theta_H)~, \cr
\noalign{\kern 5pt}
&&\hskip -1cm
m^F_{ab}(\hat \theta_H + \pi) = - m^F_{ab} (\hat \theta_H) 
= m^F_{ab}  (- \hat \theta_H) ~.
\label{effV1}
\eeqn
$m_W(0) = m_Z(0) = 0$  as the EW symmetry is recovered at $\theta_H=0$.
The set of fermion masses $ \{ - m^F_{ab} \}$ gives the same physics as 
the set $ \{  m^F_{ab} \}$ does.  The relative signs in the  equalities for $m^F_{ab}$
have been fixed by explicit evaluation at the tree level.\cite{HK}  

It has been shown in ref.\ \cite{HOOS} that $V_\eff (\theta_H)$ is minimized 
precisely at $\theta_H= \pm \onehalf \pi$ due to the contribution from  the 
top quark in the RS spacetime.  
The relations in (\ref{effV1}), then,  imply that 
all of the functions $V_\eff (\hat \theta_H)$, $m_{W, Z}^2 (\hat \theta_H)$,
and $m^F_{ab} (\hat \theta_H)$ satisfy a relation 
$F(\onehalf \pi + f_H^{-1} H) = F(\onehalf \pi - f_H^{-1} H)$.  
They are even functions of $H$ when expanded around $\theta_H = \pm \onehalf \pi$.

It follows that all odd-power Higgs couplings  $H^{2\ell +1}$,  
$H^{2\ell +1} W_\mu^\dagger W^\mu$, $H^{2\ell +1} Z_\mu  Z^\mu$, 
and $H^{2\ell +1} \psibar_a \psi_b$,  vanish.
In particular, the vanishing $WWH$ and $ZZH$ couplings  signal significant deviation 
from the standard model.  
Even if $m_H < m_W$, the LEP2 bound for the Higgs mass is evaded.
Further, Yukawa couplings  for $H \psibar_a \psi_b$ operators
vanish identically, too.

The derivative couplings such as 
$\dd_\mu \hat \theta_H \psibar \gamma^\mu \gamma^5\psi$
and $\dd_\mu \hat \theta_H Z_\nu (\dd^\mu Z^\nu - \dd^\nu Z^\mu)$
are all forbidden by the mirror reflection symmetry.
We observe that the effective interactions at low energies are invariant
under $H(x) \go - H(x)$ with all other fields kept intact
at $\theta_H = \pm \onehalf \pi$.   We call it the $H$-parity.   
Among low energy fields only the Higgs field is $H$-parity odd.
The Higgs boson becomes stable, protected by the $H$-parity conservation.
We stress that the $H$-parity has emerged dynamically, unlike in the 
models of refs.\ \cite{Barbieri, Gustafsson} where an additional Higgs 
doublet with odd parity is introduced by hand.

The mass functions are evaluated in the RS space.
It is found in refs.\ \cite{HOOS, HK, HS2} that, to a good approximation, 
\beqn
&&\hskip -1cm
m_W(\hat \theta_H) \sim \cos \theta_W m_Z(\hat \theta_H)
\sim \frac{1}{2} g f_H \sin \hat \theta_H ~, \cr
\noalign{\kern 5pt}
&&\hskip -1cm
m^F_a (\hat \theta_H) \sim \lambda_a \sin \hat \theta_H ~,
\label{mass2}
\eeqn
where $\theta_W$ is the weak mixing angle and the fermion mass matrix
has been approximated by a diagonal one $m^F_{ab} = m^F_a \delta_{ab}$.
If  a fermion belongs to  spinor representation of $SO(5)$,  
one would obtain  $m^F_a \sim  \lambda_a \sin \onehalf \hat  \theta_H$.
As $\theta_H = \onehalf \pi$, one finds that $m_W \sim \onehalf g f_H$ and 
$m^F_a \sim \lambda_a$.  The value of $f_H$ is given by $f_H \sim 246\,$GeV.
We note that this differs from the vev of the Higgs field $f_H \theta_H$.

Inserting (\ref{mass2}) into (\ref{effective1}), one finds the various Higgs
couplings;
\beeq
{\cal L}_\eff \sim - \Big\{ m_W^2 W_\mu^\dagger W^\mu 
     + \frac{1}{2} m_Z^2 Z_\mu Z^\mu \Big\}  \cos^2 \frac{H}{f_H} 
- \sum_a m_a \psibar_a \psi_a \cos \frac{H}{f_H} ~.
\label{effective2}
\eneq
The $WWHH$ coupling is given by $\frac{1}{4} g^2 W_\mu^\dagger W^\mu H^2$, 
which is $(-1)$ times the coupling in the standard 
model.\footnote{We use $\mathrm{diag.}(-+++)$ as 4D Minkowski metric.}
This coupling includes contributions coming from tree diagrams containing 
KK excited states $W_n$ of $W$ in the intermediate states with two vertices
$W W_n H$.\cite{Sakamura} The $\psibar \psi H^2$ coupling is given by
$(m_a/2 f_H^2) \psibar_a \psi_a H^2$.
It is generated by two vertices $\psi \psi_n H$ where $\psi_n$ is the $n$-th
KK excited state of $\psi$. One comment is in order.  
The approximate formula for the fermion mass function in (\ref{mass2}) 
may need corrections, depending on the details of the model.
The symmetry property leads, in general, to
$m (\hat \theta_H)= \sum_{n=0}^\infty   b_{2n+1} \sin[(2n+1)\hat\theta_H]$.
Accordingly the $\bar\psi\psi H^2$ coupling constant may be altered.

Gauge-Higgs unification models under consideration are characterized with 
two parameters $f_H$ and $m_H$ at low energies.  In a minimal model in
the RS warped space, $f_H$ is fixed around 246 GeV by 
$m_W \sim \onehalf g f_H$. 
The value of $m_H$, on the other hand, depends
on the details of the matter content in the models.  
\ignore{It was predicted around 50 GeV in ref.\ \cite{HOOS},
but can be larger with different brane fermions introduced.}
In the following numerical analysis, we fix $f_H=246\,\mathrm{GeV}$, 
whereas $m_H$ is treated as a free parameter.

With all the Higgs couplings at hand, one can estimate 
the annihilation rates of Higgs bosons in the early universe to 
determine its relic abundance as the cold dark matter.
A rough estimate may be made with the following formula:
$\Omega_H h^2\simeq 3\times 10^{-27}\,\mathrm{cm}^3\mathrm{s}^{-1}/
\langle\sigma v\rangle$, where $\Omega_H$ is the present mass density
of the Higgs boson normalized by the critical density, $h$ denotes
the Hubble constant in units of $100\,\mathrm{km}\,\mathrm{s}^{-1}\,
\mathrm{Mpc}^{-1}$, and $\langle\sigma v\rangle$ is the thermal
average of the total annihilation cross section of the Higgs bosons 
multiplied by the relative velocity. 
The present mass density of cold dark matter is determined by 
WMAP collaboration as 
$\Omega_\mathrm{CDM} h^2=0.1131\pm 0.0034.$\cite{DM2}

Suppose that the Higgs mass is sufficiently smaller than  $m_W$. 
The dominant annihilation process is $HH \go b \bar b$, and
the abundance based on the above formula turns out much
larger than the WMAP value. 
If the Higgs boson is heavier than $W$,
$HH \go W^+W^-$ dominates, and the relic abundance turns out 
much smaller than the WMAP value. Thus, we expect that
a Higgs boson slightly lighter than the W boson explains the
cold dark matter abundance observed by WMAP.

To determine a favored Higgs mass precisely, we have employed
a more elaborated formula to evaluate the relic abundance
\cite{DM3,KT,BHS}. The annihilation rate per unit particle number density
$\sigma v$ is expanded in a non-relativistic manner as 
$\sigma v=a+bv^2+O(v^4)$, and the relic abundance is given by
$\Omega_H h^2\simeq 2.82\times 10^8\,Y_\infty\,(m_H/\mathrm{GeV})$
where 
$Y_\infty^{-1}=0.264\,g_*^{1/2}m_\mathrm{pl}\,m_H\{a+3(b-a/4)/x_f\}/x_f$.
The freeze-out parameter $x_f = m_H/T_f$,  where $T_f$ is the freeze-out 
temperature,  is  determined by
$x_f=\ln [0.0382 m_\mathrm{pl}(a+6b/x_f)c(2+c)m_H/(g_* x_f)^{1/2}]$
with $c\simeq 0.5$. 

We take the following annihilation modes into account: 
$b\bar b$, $V^{(*)}V^{(*)}$,  $\tau\bar\tau$ and $c\bar c$.
Here $V$ denotes $W$ or $Z$,
and $V^*$ means a virtual gauge boson that eventually goes into
a pair of a fermion and an anti-fermion. Since we are interested in 
the threshold region of the $WW$ final state, the inclusion of 
3-body and 4-body final states via virtual gauge boson(s) is mandatory.

The cross section of the annihilation process into a final state $X$, 
$\sigma(HH\rightarrow X)$, is  obtained from the decay rate into $X$,
$\Gamma(H\rightarrow  X)$, in the standard model \cite{djouadi}, 
by replacing the relevant vertices in the standard model by
those in the present model, with an appropriate change in kinematical factors. 
For example, the cross section for $HH \go f\bar f$ is given by
\beeq
 \sigma_{f\bar f}=
  \frac{N_c}{8\pi\bar\beta_i}\,\frac{m_f^2}{f_H^4}
  \left(1-\frac{4m_f^2}{s}\right)^{3/2}  ~,
  \label{rate1}
\eneq
where $\sqrt{s}$ is the center-of-mass energy, 
$\bar\beta_i=\sqrt{1-4m_H^2/s}$, and $N_c$ is the number of
colors. The annihilation rate for $H H \go V^{(*)}V^{(*)}$ is given by
\beeq
 \sigma_{V^{(*)}V^{(*)}}=
  \frac{1}{\pi^2}
  \int_0^s dq_1^2\,D_V(q_1^2)
  \int_0^{(\sqrt{s}-\sqrt{q_1^2})^2} dq_2^2\,D_V(q_2^2)\,\sigma_0 ~,
  \label{rate2}
\eneq
where $ D_V(q^2)=m_V\Gamma_V/ \big\{ (q^2-m_V^2)^2+m_V^2\Gamma_V^2 \big\}$,
and $\sigma_0$ is the annihilation cross section into a pair of
intermediate virtual vector bosons,  given by
$ \sigma_0=
  (G_F^2 s/16\pi\bar\beta_i) \,\delta_V\lambda^{1/2}
   [\lambda + 12q_1^2 q_2^2/s^2 ] $
with $\delta_{W(Z)}=2(1)$, and 
$\lambda(q_1^2,q_2^2;s)
= s^{-2} \big\{ (s - q_1^2- q_2^2)^2 - 4q_1^2 q_2^2 \big\}$.

\begin{figure}
\begin{center}
 \includegraphics[height=6cm]{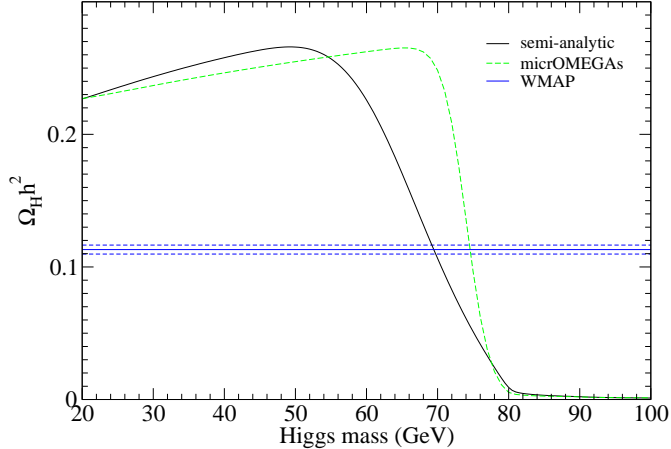} 
\caption{{\bf }  Thermal relic density of Higgs boson DM with $f_H = 246\,$GeV.
The solid (black) curve is obtained by the semi-analytic formulae, 
whereas the dashed (green) curve by the micrOMEGAs.
The horizontal band is the WMAP data 
$\Omega_{\rm CDM} h^2 = 0.1131 \pm 0.0034$.}
\label{fig:omegah2}
\end{center}
\end{figure}

Fig.~\ref{fig:omegah2} shows our numerical results for the relic 
abundance of the Higgs CDM. 
The solid (black) curve is obtained by the semi-analytic formulae
with all the aforementioned channels included in the total annihilation rate.
The horizontal (blue) lines near 0.1 indicate the allowed range by WMAP.
The favored Higgs mass is found around $70\,\mathrm{GeV}$ in a
rather narrow range. The freeze-out parameter is $x_f\sim 21$
in the favored region, which corresponds to 
$T_f\sim 3 \, \mathrm{GeV}$.

At $m_H=70\,\mathrm{GeV}$, the annhilation rates of the relevant
modes in the non-relativistic limit are 
$\sigma v=(7.3,\ 11,\ 1.5)\times 10^{-27}\mathrm{cm}^3/\mathrm{s}$ for 
$b\bar b$, $W^{(*)}W^{(*)}$, and $Z^{(*)}Z^{(*)}$ respectively. 
The $W^{(*)}W^{(*)}$ mode is larger than the $b\bar b$ mode 
even below the $WW$ threshold,  which confirms 
the importance of the $W^{(*)}W^{(*)}$ mode.
All the remaining modes including the $gg$ mode have smaller rates.

Importance of the 3-body and 4-body final states through 
$HH \rightarrow V^{(*)} V^{(*)}$ below the threshold of $VV$ pair can
be seen by comparing the solid (black) curve with 
the dashed (green) curve, which was obtained using the 
micrOMEGAs 2.2 \cite{Belanger} with two-body final states only.
The $WW$ channel opens even below the threshold due to the thermal energy.  
The micrOMEGAs including only 2-body final states gives 
$m_H\simeq 75\,\mathrm{GeV}$ as a favored Higgs mass. 
The relative contributions of
the $b\bar b$ and $WW$ modes are 34\% and 61\%,  respectively. 
The result obtained from the  semi-analytic formulae agrees with the
result from the micrOMEGAs with only 2-body final states well below 
the $WW$ threshold.  
However, near the $WW$ threshold, the relic density from 
micrOMEGAs with 2-body final states in the $HH$ annihilation yields substantially larger $\Omega_H h^2$ than the  semi-analytic treatment, although it includes thermally allowed 2-body final state 
$HH \rightarrow WW$.
Near the $WW$ threshold, it is important to include the virtual $W$ effect correctly in order to get an accurate behavior of the relic density across 
the threshold.

If Higgs bosons constitute the cold dark matter of the universe, 
they can be detected by observing Higgs-nucleon elastic scattering 
process, $HN \go HN$.
The relevant part of the effective interaction
(\ref{effective2}) is
${\cal L}_\eff=   (H^2/ 2 f_H^2) \sum_f m_f  \bar f f$.

To evaluate the direct detection rate one needs to incorporate
QCD corrections \cite{Shifman}. After integrating out the heavy
quarks, we obtain the effective Lagrangian at a hadronic scale:
\begin{equation}
{\cal L}_\eff\simeq
 \frac{H^2}{2 f_H^2}\left[\sum_{q=u,d,s}m_q\bar q q
                          -\frac{\alpha_s}{4\pi}G^a_{\mu\nu}G^{a\,\mu\nu}
                    \right]\,
\end{equation}
where $c$, $b$ and $t$ quarks are integrated out, and $G^a_{\mu\nu}$
denotes the gluon field strength. 

This effective Lagrangian leads to the following effective Higgs-nucleon
coupling:
\begin{equation}
{\cal L}_{HN} \simeq \frac{2+7f_N}{9} ~\frac{m_N}{2 f_H^2} ~H^2
\overline{N} N ~ , 
\end{equation}
where $f_N=\sum_{q=u,d,s}f^N_q$ and 
$\langle N|m_q\bar q q|N\rangle=m_N f^N_q$.  
The relation $ \la  N| (\alpha_s/8\pi) GG |N\ra = - (m_N/9) ( 1 - f_N)$
has been used.  With this coupling, the  spin-independent (SI)
Higgs-nucleon scattering cross section is evaluated to be
\beeq
\sigma_{\rm SI} \simeq
\frac{1}{4\pi} \bigg( \frac{2 + 7 f_N}{9} \bigg)^2 
\frac{m_N^4}{f_H^4 ( m_H + m_N)^2} ~,
\label{rate4}
\eneq
in the non-relativistic limit.
There is a considerable uncertainty in the value of $f_N$ stemming from 
that in the $\sigma_{\pi N}$ term. 
The value $f_N$ quoted in ref.'s\ \cite{Belanger} and \cite{Ellis:2008hf} 
are  $f_N\sim (0.31 \sim 0.41)$ and $f_N\sim (0.20 \sim 0.45)$, respectively.  
On the other hand,  
the recent lattice calculation gives a smaller value for $f_s^N$ and 
$f_N \sim 0.07$ \cite{onogi}. We choose $f_N = 0.3$ and 0.1 for 
the purpose of illustration in the following. 

\begin{figure}
\begin{center}
\includegraphics[height=6cm]{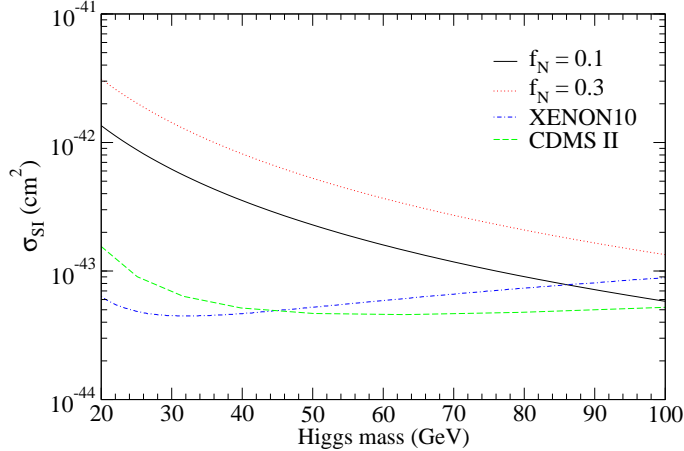} 
\caption{Spin-independent $HN$ scattering cross sections as 
functions of Higgs mass along with the bounds of CDMS II (dashed, green)  
and XENON10 (dasd-dotted, blue).
The solid (black) and the dotted (red) curves are for two different values 
of $f_N=0.1$ and $0.3$, respectively.}
\label{fig:dd_dm}
\end{center}
\end{figure}

In fig.\ \ref{fig:dd_dm}, we show the Higgs-nucleon scattering 
cross sections as functions of Higgs mass for two different values of 
$f_N=0.3$  and 0.1 with $f_H = 246$ GeV. 
The present experimental upper bounds for
the spin-independent WIMP-nucleon cross sections from 
CDMS II \cite{Ahmed:2008eu} and XENON10 \cite{Angle:2007uj}
are depicted as well. For $m_H=70\,\mathrm{GeV}$, the experimental
bound is $\sigma_\mathrm{SI} \lesssim 5 \times 10^{-44}\,\mathrm{cm}^2$ 
at 90 \% CL, whereas our prediction is  
$\sigma_{SI}\simeq (1.2 - 2.7)\times 10^{-43}\,\mathrm{cm}^2$
for $f_N = (0.1 - 0.3)$.

This does not necessarily mean that the present model is excluded.
The direct detection signals are proportional to 
the local density of cold dark matter, $\rho_0$, which has not 
been measured by experiments. In most cases including CDMS II and 
XENON10, 
the experimental bounds are derived under assuming $\rho_0 = 0.3$ GeV/cm$^3$.
For a spherical and smooth halo, $\rho_0 = 0.2 - 0.6$ GeV/cm$^3$ seems 
a reasonable range \cite{DM3,BHS}. 
Taking the lower value relaxes the constraint.
Further it has been argued  that $\rho_0$ can be as small as $0.04$ GeV/cm$^3$ 
for non-smooth distribution of dark matter in the
Galactic halo \cite{DMKK}, which makes $m_H = 70\,$GeV consistent with the data. 
On the theoretical side, as mentioned below eq.~(\ref{effective2}),
the $HH\bar f f$ coupling constants may be reduced if the mass function 
$m_f(\theta_H)$ has more general $\theta_H$-dependence. 
In this case the Higgs-nucleon cross section is decreased,
while  the Higgs relic abundance, which depends on both the $HH\bar b b$ and 
$HHWW$ couplings, 
is kept unchanged by appropriately increasing $m_H$. 
To summarize, it is premature to exclude the Higgs dark matter scenario 
based on the current bounds from CDMS II and XENON10.

Signals from pair annihilation of Higgs bosons in the Galactic halo 
into two $\gamma$'s or a $\gamma$ and a $Z$ boson 
may be seen, too. Expected signals are two (nearly) monochromatic
gamma lines of $E_\gamma= m_H(\simeq 70\mathrm{GeV})$ and 
$E_\gamma= m_H-m_Z^2/(4 m_H)(\simeq 40\mathrm{GeV})$.
Their rates are estimated to be 
$\sigma_{\gamma\gamma}\,v|_{v\rightarrow 0}
 \simeq 4.3\times 10^{-29}\mathrm{cm}^3/\mathrm{s}$,
and $\sigma_{\gamma Z}\,v|_{v\rightarrow 0}
 \simeq 5.4\times 10^{-29}\mathrm{cm}^3/\mathrm{s}$
for $m_H = 70\,\mathrm{GeV}$ and $f_H = 246\,$GeV. 
Comparing these rates with those in 
the inert doublet model \cite{Gustafsson}, we expect that such 
monochromatic gamma rays could be observed by FERMI Gamma-ray
Space Telescope (formerly GLAST). 
However the signals will be less pronounced in our model 
compared with the inert doublet model, since the annihilation cross 
section is dominated by $HH \rightarrow b \bar{b}, 
W^{(*)} W^{(*)}$ in our case,  which yield secondary 
photons with a continuum spectrum.
\ignore{, if the boost factor is $O(100)$.}

In this letter we have shown that 
the Higgs boson becomes absolutely stable in a class of  
the gauge-Higgs  unification models.  
The stability of the Higgs boson is protected by a new dynamically emerging
parity, $H$-parity.
Higgs bosons become the cold dark matter in the universe.   
The observed dark matter density is obtained with $m_H \sim 70  \,$GeV.  
Although the direct detection rate for the Higgs-nucleon elastic scattering  
is found slightly above the current upper bounds, the prediction in the 
gauge-Higgs unification model may be consistent in view of  
many uncertainties involved.
If the Higgs boson is absolutely stable, 
the way of finding Higgs bosons in collider experiments must be
scrutinized.  Higgs bosons appear as missing energies and momenta.
We shall come back to these points in future.

\vskip .5cm

\leftline{\bf Acknowledgments}
We are grateful to P. Gondolo, K. Olive, S. Scopel  and Jonghee Yoo
for useful discussions and communications.
This work was supported in part 
by  Scientific Grants from the Ministry of Education and Science, 
Grant No.\ 20244028, Grant No.\ 20025004,  and Grant No.\ 50324744 (Y.H.),
and Grant No.\ 20244037 (M.T.).


\def\jnl#1#2#3#4{{#1}{\bf #2} (#4) #3}

\def\Zphys{{\em Z.\ Phys.} }
\def\jssc{{\em J.\ Solid State Chem.\ }}
\def\jpsJ{{\em J.\ Phys.\ Soc.\ Japan }}
\def\ptps{{\em Prog.\ Theoret.\ Phys.\ Suppl.\ }}
\def\PTP{{\em Prog.\ Theoret.\ Phys.\  }}

\def\JMP{{\em J. Math.\ Phys.} }
\def\NPB{{\em Nucl.\ Phys.} B}
\def\NP{{\em Nucl.\ Phys.} }
\def\PLB{{\em Phys.\ Lett.} B}
\def\PL{{\em Phys.\ Lett.} }
\def\PRL{\em Phys.\ Rev.\ Lett. }
\def\PRB{{\em Phys.\ Rev.} B}
\def\PRD{{\em Phys.\ Rev.} D}
\def\PRe{{\em Phys.\ Rep.} }
\def\AP{{\em Ann.\ Phys.\ (N.Y.)} }
\def\RMP{{\em Rev.\ Mod.\ Phys.} }
\def\ZPC{{\em Z.\ Phys.} C}
\def\SCI{\em Science}
\def\CMP{\em Comm.\ Math.\ Phys. }
\def\MPLA{{\em Mod.\ Phys.\ Lett.} A}
\def\IJMPA{{\em Int.\ J.\ Mod.\ Phys.} A}
\def\IJMPB{{\em Int.\ J.\ Mod.\ Phys.} B}
\def\EPJC{{\em Eur.\ Phys.\ J.} C}
\def\PR{{\em Phys.\ Rev.} }
\def\JHEP{{\em JHEP} }
\def\cmp{{\em Com.\ Math.\ Phys.}}
\def\JPA{{\em J.\  Phys.} A}
\def\JPG{{\em J.\  Phys.} G}
\def\NJP{{\em New.\ J.\  Phys.} }
\def\CQG{\em Class.\ Quant.\ Grav. }
\def\ATMP{{\em Adv.\ Theoret.\ Math.\ Phys.} }
\def\ibid{{\em ibid.} }

\renewenvironment{thebibliography}[1]
         {\begin{list}{[$\,$\arabic{enumi}$\,$]}  
         {\usecounter{enumi}\setlength{\parsep}{0pt}
          \setlength{\itemsep}{0pt}  \renewcommand{\baselinestretch}{1.2}
          \settowidth
         {\labelwidth}{#1 ~ ~}\sloppy}}{\end{list}}

\def\reftitle#1{}                


\vskip 1.cm

\begin{thebibliography}{99}
\small
\baselineskip=14pt

\leftline{\bf References}

\bibitem{DM1}
M.\ Tegmark {\it et al}., 
\jnl{\PRD}{74}{123507}{2006}.

\bibitem{DM2}
E.~Komatsu {\it et al}., WMAP Collaboration, 
\jnl{\em Astrophys. J. Suppl.\,}{180}{330}{2009}.

\bibitem{DM3} 
G.\ Jungman, M.\ Kamionkowski and K.\ Griest,
\jnl{\PRe}{267}{195}{1996}.
\reftitle{Supersymmetric Dark Matter}

\bibitem{BHS}
G.~Bertone, D.~Hooper and J.~Silk, 
\jnl{\PRe}{405}{279}{2005}.
\reftitle{Particle dark matter: evidence, candidates and constraints}

\bibitem{djouadi}
A.~Djouadi,
\jnl{\PRe}{457}{1}{2008}.
\reftitle{The Anatomy of electro-weak symmetry breaking. I: The Higgs boson in the standard model}

\bibitem{YH1}
Y.\ Hosotani, \jnl{\PLB}{126}{309}{1983}.
\reftitle{Dynamical Mass Generation by Compact Extra Dimensions}


\bibitem{Davies1}
A.T.\ Davies and A.\ McLachlan,  
\jnl{\PLB}{200}{305}{1988};
\reftitle{Gauge group breaking by Wilson loops}
\jnl{\NPB}{317}{237}{1989}.
\reftitle{Congruency Class Effects in the Hosotani Model}

\bibitem{YH2}
Y.\ Hosotani, \jnl{\AP}{190}{233}{1989}.
\reftitle{Dynamics of Nonintegrable Phases and Gauge Symmetry Breaking}

\bibitem{RS}
L.\ Randall and R.\ Sundrum,  \jnl{\PRL}{83}{3370}{1999}.
\reftitle{A Large mass hierarchy from a small extra dimension}

\bibitem{HOOS} 
Y.\ Hosotani, K.\ Oda, T.\ Ohnuma  and Y.\ Sakamura, 
 \jnl{\PRD}{78}{096002}{2008},  
 {\it Erratum}, {\it ibid.} D{\bf 79} (2009) {079902(E)}.
\reftitle{Dynamical Electroweak Symmetry Breaking in SO(5)$\times$U(1) Gauge-Higgs
Unification with Top and Bottom Quarks}


\bibitem{HK}
Y.\ Hosotani and Y.\ Kobayashi, 
\jnl{\PLB}{674}{192}{2009}.  (arXiv:0812.4782[hep-ph])
\reftitle{Yukawa couplings and effective interactions in gauge-Higgs unification}

\bibitem{HS2} 
Y.\ Sakamura and Y.\ Hosotani, \jnl{\PLB}{645}{442}{2007}, 
\reftitle{WWZ, WWH and ZZH couplings in the dynamical gauge-Higgs unification
in the warped spacetime}

Y.\ Hosotani and Y.\ Sakamura, \jnl{\PTP}{118}{935}{2007}. 
\reftitle{Anomalous Higgs couplings in the $SO(5) \times U(1)$ gauge-Higgs 
unification in warped spacetime}

\bibitem{Sakamura}
Y.\ Sakamura,  \jnl{\PRD}{76}{065002}{2007}. 
\reftitle{Effective theories of gauge-Higgs unification models in warped spacetime}

\bibitem{Agashe1}
K.\ Agashe, R.\ Contino and A.\ Pomarol, 
\jnl{\NPB}{719}{165}{2005}.
\reftitle{The minimal composite Higgs model}

\bibitem{MSW}
A.D.\ Medina,  N.R.\ Shah and C.E.M.\ Wagner,
 \jnl{\PRD}{76}{095010}{2007}.    
 \reftitle{Gauge-Higgs Unification and Radiative Electroweak Symmetry 
 Breaking in Warped Extra Dimensions}

\bibitem{HHHK}
N.\ Haba, M.\ Harada, Y.\ Hosotani and Y.\ Kawamura, 
\jnl{\NPB}{657}{169}{2003};   
{\it Erratum}, {\it ibid.}  B{\bf 669} (2003) {381}.
\reftitle{Dynamical Rearrangement of Gauge Symmetry on the Orbifold $S^1/Z_2$}

\bibitem{HM}
Y.\ Hosotani and M.\ Mabe, \jnl{\PLB}{615}{257}{2005}.
\reftitle{Higgs boson mass and electroweak-gravity hierarchy
from dynamical gauge-Higgs unification in the warped spacetime}


\bibitem{Barbieri}
R.\ Barbieri, L.J.\ Hall and V.S.\ Rychkov, 
\jnl{\PRD}{74}{015007}{2006}.
\reftitle{Improved naturalness with a heavy Higgs boson: 
An alternative road to CERN LHC physics}


\bibitem{Gustafsson}
M.\ Gustafsson, E.\ Lundstr\"om, L.\ Bergstr\"om and J.\ Edsj\"o,
\jnl{\PRL}{99}{041301}{2007}.
\reftitle{Significant gamma lines from inert Higgs dark matter}


\bibitem{KT}
E.W.~Kolb and M.S.~Turner,
The Early Universe, Addison-Wesley, Redwood City, 1990.


\bibitem{Belanger}
G.\ Belanger, F.\ Boudjema, A.\ Pukhov and A.\ Semenov,
\jnl{\em Comput.~Phys.~Commun.~}{180}{747}{2009}. 
\reftitle{Dark matter direct detection rate in a generic model with
micrOMEGAs_2.2}


\bibitem{Shifman}
M.A.\ Shifman,  A.I.\ Vainshtein and V.I.\ Zakharov, 
\jnl{\PLB}{78}{443}{1978}.
\reftitle{Remarks on Higgs-boson interactions with nucleons}

\bibitem{Ellis:2008hf}
  J.~R.~Ellis, K.~A.~Olive and C.~Savage,
\reftitle{Hadronic Uncertainties in the Elastic Scattering of 
Supersymmetric Dark Matter}
  \jnl{\PRD}{77}{065026}{2008}. 

\bibitem{onogi}
H.\ Ohki {\it et al.},
\jnl{\PRD}{78}{054502}{2008}
\reftitle{Nucleon sigma term and strange quark content from lattice QCD with exact chiral symmetry}
  Phys.\ Rev.\  D {\bf 78}, 054502 (2008).




\bibitem{Ahmed:2008eu}
  Z.~Ahmed {\it et al.}  [CDMS Collaboration],
\reftitle{Search for Weakly Interacting Massive Particles with the 
First Five-Tower}
\jnl{\PRL}{102}{011301}{2009}.

\bibitem{Angle:2007uj}
  J.~Angle {\it et al.}  [XENON Collaboration],
\reftitle{First Results from the XENON10 Dark Matter Experiment 
at the Gran Sasso National Laboratory}
\jnl{\PRL}{100}{021303}{2008}. 
  

\bibitem{DMKK}
M.~Kamionkowski and S.M.~Koushiappas,
\reftitle{Galactic substructure and direct detection of dark matter}
\jnl{\PRD}{77}{103509}{2008}.



\end{thebibliography}
\end{document}